\documentclass[12pt]{article}
\usepackage{palatino,epsfig,makeidx,amsfonts,amsmath}
\usepackage[round]{natbib}

\def\E{\mbox{E}}

\def\V{\mbox{V}}

\def\N{\mbox{N}}
\def\Ga{\mbox{Ga}}
\def\GG{\mbox{GG}}

\def\Be{\mbox{Be}}

\def\d{\mbox{d}}

\def\Ex{\mbox{Ex}}

\def\SE{\mbox{SE}}

\newtheorem{theorem}{Theorem}
\newtheorem{prop}{Proposition}

\begin{document}

\title{Hierarchical sparsity priors for regression models}

\author{J.E.~Griffin and P.J.~Brown\\
School of Mathematics, Statistics and Actuarial Science,\\ University of Kent, Canterbury CT2 7NF, UK}

\maketitle

\abstract{
We focus on the increasingly important area of sparse regression problems where there are many variables and the effects of a large subset of these are negligible.  This paper describes the construction of hierarchical prior distributions when the effects are considered related. These priors allow dependence between the regression coefficients and encourage related  shrinkage towards zero of different regression coefficients. The properties of these priors are discussed and applications to linear models with interactions and generalized additive models are used as illustrations. Ideas of heredity relating different levels of interaction are encompassed 
}

\section{Introduction}

Regression modelling is an important means of understanding the effect of predictor variables on a response. These effects can be hard to estimate and interpret if the predictor variables are highly correlated (the problem of collinearity) or there are large numbers of predictor variables. These aspects are often addressed by assuming that effects are sparse (meaning that only 
a subset  of the predictor variables has a large  effect on the response). This, in turn, can lead to more interpretable models and better out-of-sample prediction.
 In a Bayesian framework, sparsity can be achieved using ``spike-and-slab'' priors \citep{mibe88} or more recently various regularization methods \citep{paca08, gribro10, caposc10, armduncly11, armdunlee12} where shrinkage of regression coefficients to, or close to, zero is encouraged.

Most work in the area of Bayesian regularization has not explicitly included any known relationships between the predictor variables in the analysis with regression coefficients considered independent {\it a priori}. However, in many data sets, there are known relationships between the predictor variables which we wish to include in the analysis. For example, suppose that we use a linear model with main effects and two-way interaction terms. One commonly used heuristic in variable selection is that a two-way interaction term can only be included if both main effects terms are included. This assumption is open to criticism and we provide a more robust implementation which allows the data to contradict the assertion. In a Bayesian framework, this heuristic can be interpreted as a belief that the absolute size of the two-way interaction coefficient is related to the two associated main effect coefficients (if either main effect has a small absolute coefficient then the interaction term must also have a small absolute coefficient). Of course, other assumptions could be made but it is clear that it is often natural to assume a relationship between the usefulness of the interaction term and the usefulness of the main effects.  By incorporating this into the prior it is left open to the data to refute this.

Several approaches have been developed in the literature to allow various relationships to be included in the analysis. 
%Perhaps, the most popular is the group lasso \citep{YuanLin06} where predictor variables are divided into groups and a %penalty function is developed for which the penalized maximum likelihood estimates of the regression coefficients in a group %are either all zero or all non-zero. Therefore, variable selection occurs at the level of the groups rather than the individual %predictor variables. A similar approach for non-overlapping groups was developed by \cite{jaobve09}. The group lasso %approach was extended to linear models with two-way interactions by \cite{YJL07} and to more complicated problems by %\cite{YJZ09}.
Prior distributions which include known relationships between variables have also been considered. A Bayesian version of the group Lasso \citep{YuanLin06} was developed by \cite{Kyung10} and \cite{Raman10}. A  different approach is taken by \cite{gribro10b} who defined priors which allow correlation between the effects rather than dependence through the absolute effect sizes (as implied by the group Lasso). It has also been applied to unifying and robustifying ridge and g-priors for regression in \cite{gribro13}.
The variable selection problem in the linear model with interactions has been approached by \cite{CHW97} using ``spike-and-slab'' prior distributions. More recently, structured priors have been proposed in biological application, {\it e.g.} \cite{StChTaVa11} and \cite{LiZh10}.

In this paper, we develop  a method for building  prior distributions for structured regression problems (where relationships between the predictor variables can be assumed).
The prior involves organising the regression coefficients in a hierarchical structure where the regression coefficients at one level depend on a subset of the effect sizes at lower levels and where the effects are less likely to be important at higher levels. This is a fairly general structure which can include different group structures
\citep[see {\it e.g.}][]{YuanLin06,jaobve09}
 in a simple way, whilst also expressing much more complicated structures. 
The methodology gives a general and relatively simple way of controlling complexity at different levels of a hierarchy through a development of sparsity.

The paper is organized as follows. Section \ref{se:priors} explains  the use of normal-gamma and normal gamma-gamma (or generalized beta mixture) priors for sparse regression problems.  Section \ref{se:hierarch} develops hierarchical  structured regression models using a hierarchical prior and uses  the linear model  with interaction terms and the  generalized additive model as motivating examples.
%and  the GAM with main effects and two-way interactions. 
The general construction and its use in specific modelling contexts is given in section \ref{se:gen_con} with properties of the priors are discussed in section \ref{se:props}. Section \ref{se:comp} briefly describes computational strategies for models using these priors. Section \ref{se:example} includes applications of the models introduced in 
section \ref{se:hierarch}. A  discussion follows in section \ref{se:disc} and 
proofs of the theorems are given in the Appendix. Shrinkage characterisation and a further example are provided in the supplementary material.

\section{Continuous priors for sparse regression}
\label{se:priors}

The normal linear regression model for an $(n\times 1)$-dimensional vector of responses $y$ and an $(n\times p)$-dimensional design matrix ${\bf X}$  is
\begin{equation}
y = \alpha{\bf} 1 + {\bf X}{\bf \beta} + {\bf \epsilon}
\label{basic_reg}
\end{equation}
where ${\bf \epsilon}\sim\N(0,\sigma^2 {\bf I}_n)$, $1$ is a $(n\times 1)$-dimensional vector of 1's, $\alpha$ is an intercept and $\beta$ is a $(p\times 1)$-dimensional vector of regression coefficients. The prior for $\alpha$ and $\sigma^{-2}$ is chosen to be  the scale-invariant choice $p(\alpha,\sigma^2)\propto \sigma^{-2}$. We will concentrate on the choice of prior for the regression coefficients $\beta$, which will be assumed independent of $\alpha$ and $\sigma^2$, in the rest of the paper. we assume that the variables have been measured on comparable scales (or scaled to have comparable scales).

Zero-mean scale mixtures of normals are a wide class of priors for regression coefficients \citep[see {\it e.g.}][]{polsco11} in which the prior density can be expressed as  
\[
\pi(\beta_i)=\int \N(0,\Psi_i)\,dG(\Psi_i)
\]
where $G$ is a distribution function with density $g$ (if it exists). 
Many priors fit into this class including the ``spike-and-slab'' prior \citep{mibe88} and stochastic search variable selection prior \citep{geomcc93}  where $G$ is chosen to  be a discrete mixing distribution with two possible values.
Alternatively, many priors use an absolutely continuous $G$ including the double exponential \citep{paca08, hans09} (leading to the Bayesian Lasso), the normal-gamma \citep{cado08, gribro10} the Bayesian elastic net \citep{hans11}, the horseshoe prior \citep{caposc10},  the normal-exponential-gamma (NEG) \citep{gribro11}, the generalized Beta mixtures \citep{armduncly11}, the generalized $t$ \citep{lecadoho12} or double Pareto prior \citep{armdunlee12} and the exponential power prior \citep{poscwi13}.

 In this paper, we will consider two priors. The normal-gamma prior \citep{cado08, gribro10} which has the form
\[
\beta_j\sim\N(0,\Psi_j),\qquad \Psi_j\sim\Ga(\lambda,\gamma).
\]
The prior variance is $\V[\beta_j]=\E(\Psi_j)=\frac{\lambda}{\gamma}$. The generalized beta mixture prior distribution \citep{armduncly11} can be expressed as a hierarchical extension of the normal-gamma prior
\[
\beta_j\sim\N(0,\Psi_j),\qquad \Psi_j\sim\Ga(\lambda,\gamma_j),\qquad \gamma_j\sim\Ga(c,d).
\]
  and the prior variance is 
$\V[\beta_j]=\frac{\lambda d}{c-1}$  if $c>1$. We will refer to this distribution as  the normal-gamma-gamma prior distribution to emphasize the link to the normal-gamma prior.
The hyperparameters have simple interpretations: $d$ is a scale parameter, $\lambda$ controls the behaviour of the distribution close to zero and $c$ controls the tail behaviour of the distribution.
The marginal density of $\beta_j$  is not available in closed form but the marginal distribution of $\Psi_j$ is a gamma-gamma distribution  which has the density
\[
g(\Psi_j)=\left(\frac{1}{d}\right)^{\lambda}\frac{\Gamma(\lambda+c)}{\Gamma(\lambda)\Gamma(c)}
\Psi_j^{\lambda-1}\left(1+\frac{\Psi_j}{d}\right)^{-(\lambda+c)}.
\]
This prior will be written $\Psi_j\sim\mbox{GG}(\lambda,c,d); $ and corresponds to the inverted-beta-2 distribution of \citet[][section 7.4.2]{raisch61}. The authors showed that the monotone transformation $\frac{\Psi_j}{\Psi_j+d}$ has  a beta distribution with parameters $\lambda$ and $c$ implying that the median of $\Psi_j$ is $d$ if $\lambda =c$. This is  a useful characterisation if  $c\le1$ and the mean does not exist.  In particular, this is true
 for the horseshoe  prior which occurs if $\lambda=c=1/2$. 
Several of the absolutely continuous priors for  regression coefficients described in Section 1 can be written as special cases of the normal-gamma-gamma distribution including the NEG distribution which arises when $\lambda=1$ and the normal-gamma  distribution which arises if $c/d=\mu$ as $c\rightarrow\infty$.

Shrinkage results for regression models which express  the posterior expectation and variance in terms of the least squares estimate of $\beta$ and the variance of its sampling distribution (for $n>p$)
have been derived by several authors including
\cite{caposc10}, \cite{gribro10} and \cite{polsco12} and illustrate how aggressively different priors will shrink regression coefficients to zero.
The sparsity of a set of regression coefficients can be considered to be the proportion which  have values close to zero. Smaller values of $\lambda$ in the normal gamma and normal gamma-gamma will increasingly favour sparser sets of regression coefficients since small coefficients are likely to be shrunk very close to zero. 
 This is intuitively reasonable since this parameter controls the shape of the distribution of $\Psi_i$ at small values for both priors, gamma and gamma-gamma. Consequently, we define the sparsity shape parameter for a prior distribution in terms of the prior density of $\Psi_i$ as $$\sup\left\{
z\left\vert \frac{p(\Psi_i)}{\Psi_i^{z-1}}\rightarrow \kappa\mbox{ as }\Psi_i\rightarrow 0\mbox{ for finite } \kappa
\right.\right\}$$ where $p(\Psi_i)$ is the prior density of $\Psi_i$. This will be simply  $\lambda$ in the case of both the normal-gamma and normal-gamma-gamma prior distributions and indicates the shape of the prior distribution of $\Psi_i$ close to zero. The use of the supremum or least upper bound leads to clearer results in some special cases discussed in section 3.2.

\section{Hierarchical sparsity priors}
\label{se:hierarch}

\subsection{Motivating Examples}
\label{sse:motivating}

Before developing our general hierarchical prior, it is useful to set the context by considering two statistical models: the linear models with interactions and the generalized additive model. These illustrate the need for priors which can express relationships between regression coefficients with different levels of sparsity for some regression coefficients.

\subsubsection{Linear models with interaction terms}
\label{sse:interact}

Variable selection and regularization methods for linear models with interactions have received some attention in the literature
 \citep{CHW97, YJL07}. The model assumes that response $y_i$ which is observed with covariates $X_{i1},\dots,X_{ip}$ can be expressed as 
\[
y_i =\alpha + \sum_{j=1}^p X_{ij}\beta_j + \sum_{j=1}^p \sum_{k=1}^{j-1} X_{ij}X_{ik}\delta_{jk} + \epsilon_i,\qquad 
\mbox{for } i=1,\dots,n
\]
where $\epsilon_i\sim\N(0,\sigma^2)$. 
It is often considered natural to make the inclusion of an interaction contingent on the inclusion of main effects.
\citet{CHW97} formalize this idea using two forms of the heredity principle.
 {\it Strong heredity} states that an interaction can only be included if both main effects are included. {\it Weak heredity}  states that an interaction can be included if at least one main effects is included.
The use of strong or weak heredity suggests beliefs which are inconsistent with an assumption of prior independence between the regression coefficients. It is also clear that, {\it a priori}, the scale of the interaction coefficient should depend on the magnitude, but not the sign, of the main effect coefficients with the coefficients of the interactions being sparser than the coefficients of the main effects.

%  $\lambda_1$, the marginal sparsity of the interactions is $\min\{\lambda_1,\lambda_2\}$, and the conditional sparsity of the interactions is $\lambda_2$.

 %  The results in section 4 suggest that the marginal sparsity of the main effects is $\lambda_1$, the marginal sparsity of the interactions is $\min\{\lambda_1,\lambda_2\}$, and the conditional sparsity of the interactions is $\lambda_2$.
 
 %In either case, in  terms of a standardised scale as allowed by the proposition in the Appendix A, we assume the variance parameters are independently given as $\eta^{(1)}_j\sim%\GG(\lambda_1,c,1)$ and $\eta^{(2)}_j\sim\GG(\lambda_2,c,1)$. The results in section 4 suggest that the marginal sparsity of the main effects is $\lambda_1$, the marginal sparsity %of the interactions is $\min\{\lambda_1,\lambda_2\}$, and the conditional sparsity of the interactions is $\lambda_2$.
% It seems sensible to generally assume that the interactions are sparser than the main effects which is consistent with $\lambda_2<\lambda_1$. 

\subsubsection{Generalized additive models}
\label{sse:GAM}

The generalized additive model (GAM) \citep{hastib93} is a non-linear regression model which represents the mean of the response as a linear combination of potentially non-linear functions of each variable so that
\[
y_i=\sum_{j=1}^p f_j(X_{ij})+\epsilon_i
\]
where $\epsilon_i\sim\N(0,\sigma^2)$ and $f_j$ are function to be estimated from the data. Reviews of Bayesian analysis of these models are given by \cite{KohnSmithChan:01}
and
\cite{DeHoMaSm02}. A common approach assumes that each non-linear function can be represented as a linear combination of basis functions so that, {\it e.g.},
\[
f_j(X_{ij}) =  \theta_{j} X_{ij}^k+\sum_{k=1}^K \gamma_{jk} \,g(X_{ij},\tau_{jk})
\]
where  $g(x,\tau_{j1}),\dots,g(x,\tau_{jK})$ are a set of basis functions with knot points $\tau_{j1}, \dots, \tau_{jK}.$ This leads to a linear model for the responses
\[
y_i=\sum_{j=1}^p f_j(X_{ij})+\epsilon_i
=\sum_{j=1}^p 
 \theta_{j} X_{ij}+\sum_{j=1}^p\sum_{k=1}^{K} \gamma_{jk} \,g(X_{ij},\tau_{jk})+\epsilon_i.
\]
The set of knot points is often chosen to be relatively large and many $\gamma_{jk}$'s are set to zero to avoid over-fitting. In a Bayesian framework, this is usually approached as a variable selection problem and so we effectively have $p$ different variable selection problems (one for each variable). We will refer to this as selection at the {\it basis level}.
 There is also the more standard variable selection problem of choosing a subset of the variables which are useful for predicting the response. The effect of the $j$-th variable is removed from the model if $\theta_{j}$ and $\gamma_{j1},\dots,\gamma_{jK}$ are all set to zero. We refer to this as selection at the {\it variable level}.
In this model, prior independence between the coefficients for the $j$-th variable $(\theta_{j})$ and $(\gamma_{j1},\dots,\gamma_{jK})$ seems unreasonable and dependence in size (rather than the sign) of these coefficients will be reasonable in many problems. 
Typically, we would like different amounts of sparsity at the basis level and the variable level which suggests a prior with at least two sparsity parameters.

\subsection{General construction}
\label{se:gen_con}

The examples in section \ref{sse:motivating} illustrate the need for priors which allow dependence in the size of regression coefficients but not their sign with hyperparameters that control the amount of sparsity implied by the prior for different regression coefficients.
The Bayesian group lasso \citep{Kyung10, Raman10} is one example of a prior which allows dependence between the size of regression coefficients but no correlation in the signs. It is assumed that the regression coefficients are divided into disjoint groups $b_1,b_2,\dots,b_G$ where $b_i$ is the $(p_i\times 1)$-dimensional vector of regression coefficients for the $i$-th group. The prior assumes that $b_1,b_2,\dots,b_G$ are independent and $b_i\sim\N\left(0,\Psi_i D^{(i)}\right)$ where $\Psi_i$ is given a gamma prior distribution and $D^{(i)}$ is a $(p_i\times p_i)$-dimensional matrix. This induces correlation in the conditional variances of the regression coefficients,  $\Psi_iD^{(i)}_{jj}$ for $j=1,\dots,p_i$, but not necessarily in the regression coefficients (the correlation between $b_{ij}$ and $b_{ik}$ will be zero if $D^{(i)}_{jk}=0$).

The group lasso prior is a simple way of building dependence between regression coefficients if they can be divided into groups. We consider a more general structure for the prior of the regression coefficients, $\beta=(\beta_1,\dots,\beta_p)$,
in (\ref{basic_reg}). 
 We assume that the elements of 
$\beta$ are independent conditional on $\Psi=(\Psi_1,\dots,\Psi_p)$ and 
\[
\beta_j \sim \N(0,\Psi_j),\qquad j=1,\dots,p.
\]
The parameter $\Psi_j$ is the conditional variance of $\beta_j$ and smaller values of $\Psi_j$ imply typically smaller values of  $\vert\beta_j\vert$. Building hierarchical priors for $\Psi$ allows the construction of a prior with correlated $\Psi$ but not  $\beta$. This form of dependence is important.
The scale $\Psi_j$ can be interpreted as the importance of the $j$-th variable in the regression and so correlating $\Psi_j$ and $\Psi_k$ implies a relationship between the importance of the $j$-th and $k$-th variables. Lack of correlation between the regression coefficient imples, for example, no correlation in the sign of regression coefficients, which is a natural assumption in many regression problems. The construction could be extended to a prior where the regression coefficients are correlated by assuming that $\beta$ are dependent conditional on $\Psi$ but this is not considered in this paper.

%The main results for shrinkage are developed in the next section relative to both normal-gamma and normal-gamma-gamma %priors.

%Most work on  Bayesian regularization methods has concentrated on priors where the regression coefficients are %independent.  \cite{gribro10b} propose the correlated normal-gamma (CNG) prior distributions under which the regression %coefficients are linear combinations of independent normal-gamma random variables and so are correlated.

We assume that  the regression coefficients can be arranged in levels. Usually, the first level will refer to a linear regression with main effects only and later levels will add additional flexibility (and complexity) to the model ({\it e.g.} all interactions). Typically, we would assume the regression coefficients become sparser at higher levels.
In general, let there be $L$ levels and
 $\beta^{(l)}$ be the $(p_l\times 1)$-dimensional vector of regression coefficients in the $l$-th level. The regression coefficients at a particular level will have the same  sparsity {\it a priori} and their scales will usually depend on scales of regression coefficients in lower levels. Our general prior assumes that
 \[
\beta^{(l)}_j\stackrel{i.i.d.}{\sim}\N\left(0,\Psi^{(l)}_j\right),\qquad 
j=1,\dots,p_l,\quad l=1,\dots,L,
\]
and
\begin{equation}
\Psi^{(l)}_j = s_j^{(l)}\,d\,\frac{f_{jl}\left(\Psi^{(1)},\dots,\Psi^{(l-1)}\right)}{\E[f_{jl}\left(\Psi^{(1)},\dots,\Psi^{(l-1)}\right)]}
\,\eta^{(l)}_j, \qquad 
j=1,\dots,p_l, \quad l=1,\dots,L
\label{def_psi}
\end{equation}
where $\eta^{(l)}_j$ are given independent prior distributions with mean 1 and $s_j^{(l)}$ is the  sparsity shape parameter of $\eta_j^{(l)}$. It follows that $\E\left[\Psi_j^{(l)}\right] = s_j^{(l)}\, d$ which mimics the normal-gamma prior distribution where the sparsity shape parameter is the shape parameter of the gamma distributions and $d$ can be interpreted as a scale parameter.  The Bayesian group lasso arises from taking a single level, setting $\Psi_i^{(1)}=\Psi_j^{(1)}$ if $i$ and $j$ are in the same group and choosing $\eta_j^{(l)}$ to have a gamma distribution.

The function $f_{jl}$ will usually be a simple function using combinations of additions and multiplications to allow easy calculation of its expectation and clear understanding of the sparsity. 
Products have the useful property of being small if one element in the product is small and sum have the useful property 
of being small if all elements in the sum are small.
Other choices of $f_{jl},$  such as minimum or maximum are possible, but would not lead to such simple calculation and interpretation.

\subsubsection{Linear model with interaction terms}
\label{se:interact_prior}

 In our framework, we interpret {\it strong heredity} as a prior belief that $\delta_{jk}$ will be strongly shrunk to zero if either $\beta_j$ or $\beta_k$ are strongly shrunk to zero. 
We interpret {\it weak heredity} as a prior belief that $\delta_{jk}$ will be strongly shrunk to zero if both $\beta_j$ and $\beta_k$ are strongly shrunk to zero. 
 These prior beliefs can be represented using a hierarchical sparsity prior. First, we define two levels: the interaction level and the main effect level. The first level (the main effect level) has $p_1=p$ terms listed as $\beta_1,\dots,\beta_p$ and the 
second level
(the interaction level) has $p_2=p(p-1)/2$ terms listed as  $\delta_{jk}$ for $k=1,\dots,j-1$, $j=1,\dots,p$. 
In the case of strong heredity, we use the prior
\[
\beta_j\sim\N\left(0,\lambda_1\,d\,\eta^{(1)}_j\right),\qquad
\eta^{(1)}_j\sim \GG\left(\lambda_1,c,\frac{c-1}{\lambda_1}\right),
\] 
\[
 \delta_{jk}\sim\N\left(0,\lambda_2\,d\,\eta^{(2)}_{jk}\,\eta^{(1)}_j\,\eta^{(1)}_k\right),\mbox{ and }
\eta^{(2)}_{jk}\sim \frac{c}{\lambda_2}\,\GG\left(\lambda_2,c,\frac{c-1}{\lambda_2}\right),
\]
The prior variance of $\delta_{jk}$ is small if at least one of , $\eta_j^{(1)}$, $\eta_k^{(1)}$ (and hence also the prior variances of  $\beta_j$ and $\beta_k$) or $\eta_{jk}^{(2)}$ is small . Therefore, an interaction term $\delta_{jk}$ will tend to be small (since its variance is small) if either $\eta^{(2)}_{jk}$ is small or if at least one of $\beta_j$ or $\beta_k$ are small (which implies that its prior variance is small). 
In the case of weak heredity, we use the prior
\[
\beta_j\sim\N\left(0,\lambda_1\,d\,\eta^{12)}_j\right),\qquad
\eta^{(1)}_j\sim \GG\left(\lambda_1,c,\frac{c-1}{\lambda_1}\right),
\] 
\[
 \delta_{jk}\sim\N\left(0,\lambda_2\,d\,\eta^{(2)}_{jk}\,\frac{1}{2}\left(\eta^{(1)}_j+\eta^{(1)}_k\right)\right),
\mbox{and } \eta^{(2)}_{jk}\sim \frac{c}{\lambda_2}\,\GG\left(\lambda_2,c,\frac{c-1}{\lambda_2}\right).
\]
The prior variance of $\delta_{jk}$ is  small if $\eta^{(2)}_{jk}$ is small or the prior variances of 
{\it both} $\beta_j$ and $\beta_k$ are small. Therefore, the interaction terms will tend to be small  if $\eta^{(2)}_{jk}$ is small or if both $\beta_j$ and $\beta_k$ are small (using similar reasoning to the strong heredity case).

\subsubsection{GAM models}
\label{se:GAM_prior}

In section~\ref{sse:GAM}, we discussed how inference in the GAM model could be seen as a
 two-level variable selection problem (at the basis level and at the variable level).
This can be approached using a hierarchical sparsity prior by defining 
the  first level (the variable level) by  $p_1=p$  terms  $\theta_{j}$ for $j=1,\dots,p$ and 
the second level (the basis level) by $p_2=pK$ terms 
$\gamma_{jk}$ for $j=1,\dots,p$,  $k=1,\dots,K$. We propose the prior
\[
\theta_{j}\sim\N\left(0,\lambda_1\,d\,\eta^{(1)}_j\right),\qquad
\eta^{(1)}_j\sim\GG\left(\lambda_1,c,\frac{c-1}{\lambda_1}\right),
\]
\[
 \gamma_{jk}\sim\N\left(0,\lambda_{2,j}\,d\,\eta^{(2)}_{jk}\,\eta^{(1)}_j\right),\mbox{ and }
\eta^{(2)}_{jk}\sim\GG\left(\lambda_{2,j},c,\frac{c-1}{\lambda_{2,j}}\right),
\]
A small value of the parameter $\eta^{(1)}_j$ implies that the $j$-th variable is unimportant and will effect the shrinkage of both the linear effect $\theta_{j}$ and basis function coefficients $\gamma_{j1},\dots,\gamma_{jK}$ leading to shrinkage at the variable level. The variable selection problem at the basis level is achieved through the different values of $\eta^{(2)}_{jk}$ which allow some basis function coefficients to be set very close to zero. The prior allows different sparsity levels for the basis function coefficients for each variable ({\it i.e.} sparsity $\lambda_{2,j}$ for the $j$-th variable).

\subsection{Properties of the prior}
\label{se:props}

A sensible choice of sparsity is essential to good estimation of the regression coefficients. Therefore, it is important to consider the sparsity shape parameters 
of the distributions of the regression coefficients
induced by 
 this prior.
The sparsity within the $l$-th level is controlled by the sparsity shape parameter of the 
marginal distribution of $\Psi_j^{(l)}$. It is also interesting to consider the sparsity shape parameter of the distribution of $\Psi_j^{(l)}$ conditional on $\Psi^{(1)},\dots,\Psi^{(l-1)}$. 
%This shows how the relationships between the scales will affect inference. 
We refer to the sparsity shape parameter of the marginal distribution of $\Psi_j^{(l)}$ as the {\it marginal sparsity shape parameter} and the sparsity shape parameter of the conditional distribution of 
$\Psi_j^{(l)}$ given $\Psi^{(1)},\dots,\Psi^{(l-1)}$ as the {\it conditional sparsity shape parameter}. We similarly distinguish between the shrinkage induced by the marginal and conditional distributions.

The conditional sparsity shape parameter and shrinkage are more easily understood than the marginal sparsity shape parameter and shrinkage. The conditional sparsity shape parameter is given by the 
sparsity shape parameter of $\eta_j^{(l)}$ and the conditional shrinkage has scale of 
$d\frac{f_{jl}\left(\Psi^{(1)},\dots,\Psi^{(l-1)}\right)}{\E[f_{jl}\left(\Psi^{(1)},\dots,\Psi^{(l-1)}\right)]}$. Therefore, smaller values of  $f_{jl}\left(\Psi^{(1)},\dots,\Psi^{(l-1)}\right)$ lead to larger amounts of shrinkage as seen by the characterisation later in  this section in Proposition 1.
To characterise the marginal sparsity shape parameter, we will consider  functions, $f_{jl}\left(\Psi^{(1)},\dots,\Psi^{(l-1)}\right)$, formed through products or sums. The use of products to define a sequence of priors with increasing shrinkage has been considered by \cite{BhDu11} in the context of factor models.

\begin{theorem}[Gamma case]
Suppose that $\eta_i \sim\Ga(\lambda_i, 1)$ for $i=1,2,\dots,K$ then
\begin{enumerate}
\item the sparsity shape parameter of $\Psi$ is $\min\{\lambda_i\}$
if  $\Psi = \prod_{i=1}^K \eta_i$. 
\item the sparsity shape parameter of  $\Psi$ is $\sum_{i=1}^K \lambda_i$
if $\Psi = \sum_{i=1}^K \eta_i$.
\end{enumerate}
\end{theorem}

An interesting special case is the product of two gamma random variables $\Psi=\eta_1\eta_2$ for which the density has the analytic expression,
\[
g(\Psi)=\frac{2}{\Gamma(\lambda_1)\Gamma(\lambda_2)}
\Psi^{(\lambda_1+\lambda_2)/2-1}K_{\vert \lambda_1-\lambda_2\vert}(2\sqrt{\Psi})
\]
where $K_{\nu}(\cdot)$ is the modified Bessel function of the third kind \citep[][pg. 374]{abrste64}.
The distribution is referred to as the $K$-distribution \citep{Jakpus78} in several areas of physics.
Using a small value approximation \citep[][eqn 9.6.9]{abrste64}, this  density at a value of $\Psi$ near zero is approximately proportional to
\[
\frac{\Gamma(\vert \lambda_1 - \lambda_2\vert)}{\Gamma(\lambda_1)\Gamma(\lambda_2)}
\Psi^{\min\{\lambda_1,\lambda_2\}-1},
\]
and so the sparsity shape parameter is $\min\{\lambda_1,\lambda_2\}$ which is 
in agreement with Theorem 1.
\begin{figure}[h!]
\begin{center}
\includegraphics[scale=1, clip, trim = 10mm 0mm 60mm 230mm]{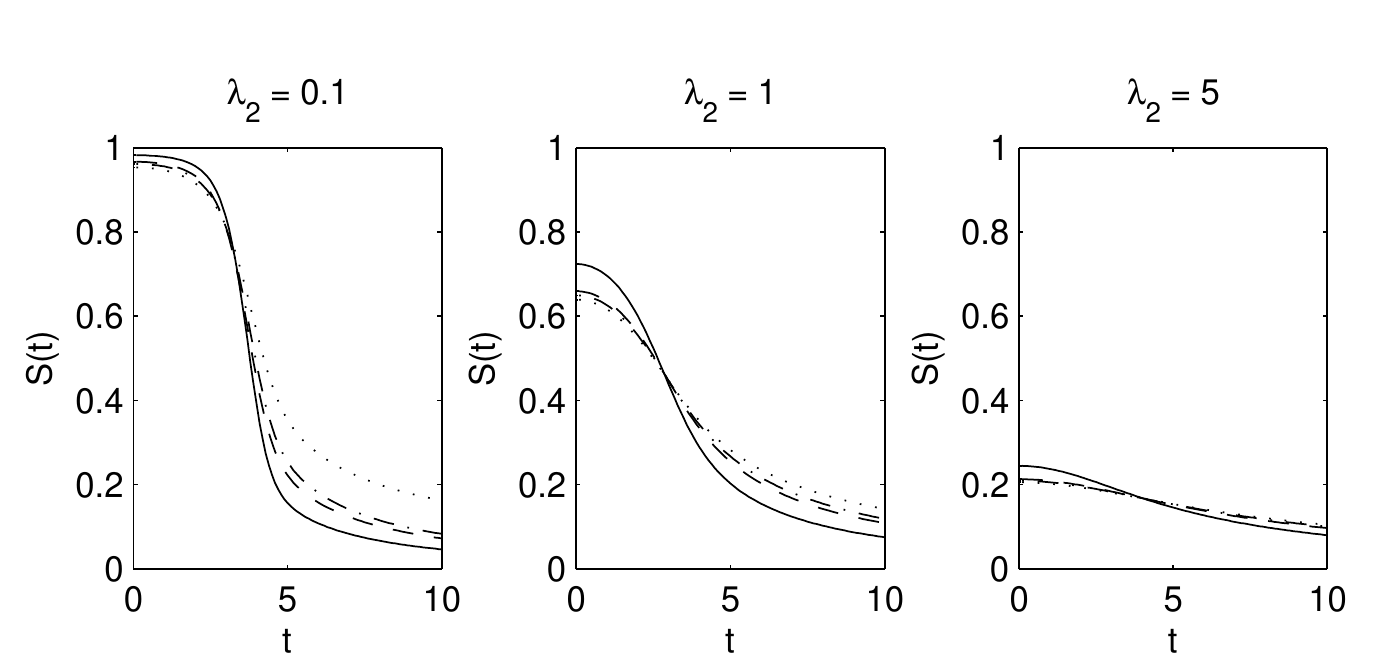}
\end{center}
\caption{\small Shrinkage profiles for various choices of products of two normal-gamma prior distribution with:
 $\lambda_1=\lambda_2$ (solid line), $\lambda_1 = 5\lambda_2$ (dashed line),  $\lambda_1 = 10\lambda_2$ (dot-dashed line) with $d=1/\SE^2$ compared to a normal-gamma with shape $\lambda_1$ (dotted line) with $d=1/\SE^2$. 
}\label{fig:comp_1}
\end{figure}
%\begin{figure}[h!]
%\begin{center}
%\includegraphics[scale=1, clip, trim = 10mm 0mm 60mm 220mm]{NG3_prod} 
%\caption{\small Shrinkage profiles for various choices of product of three normal-gamma prior distributions where %$\lambda_3=\lambda_2$ with:   $\lambda_2=\lambda_1$ (solid line),
% $\lambda_2 = 5\lambda_1$ (dashed line), 
%$\lambda_2 = 10\lambda_1$
%(dot-dashed line) 
%with $d=1/\SE^2$
%compared with
%a normal-gamma with shape $\lambda_1$ (dotted line) with $d=1/\SE^2$.
%}\label{fig:comp_2}
%\end{center}
%\end{figure}
Theorem 1 can be extended to the gamma-gamma case giving:
\begin{theorem}[Gamma-gamma case]
Suppose that $\eta_i \sim\mbox{GG}(\lambda_i, c_i,1)$ for $i=1,2,\dots,K$ then
\begin{enumerate}
\item the sparsity shape parameter of $\Psi$ is $\min\{\lambda_i\}$
if $\Psi = \prod_{i=1}^K \eta_i$.
\item the sparsity shape parameter of $\Psi$ is $\sum_{i=1}^K \lambda_i$
if $\Psi = \sum_{i=1}^K \eta_i$. 
\end{enumerate}
\end{theorem}
Therefore, the shape close to zero of the products of either a normal-gamma or normal-gamma-gamma distribution is controlled by the shape parameters $\lambda_1,\dots,\lambda_K$ rather than the other parameters.

In order to simplify presentation of shrinkage graphs across different sampling setups and  priors we can standardise the comparison.  To illustrate the method we take the single regression parameter special case of the Proposition in \citep{gribro10}.

\begin{prop}

Suppose that we have the regression model in (\ref{basic_reg})
 with a single regressor which has been centred. The intercept is given the vague prior $p(\alpha)\propto 1$ and $\beta$ has prior $\pi_{\beta}(\beta)$. Let $\tau=\beta/\mbox{SE}$
  and with t-statistic $t=\hat\beta/\SE$ where $\hat\beta$ is the least squares estimate of the regression coefficient and $\SE$ is its standard error and $\pi_{\beta}=[1/\kappa]\pi_S(\beta/\kappa)$ where $\pi_S(.)$ is a standardised version of the prior with say an interquartile range of unity.  Then
\begin{equation}
\E[\beta\vert \hat\beta]=(1-S(t))\hat\beta \label{shrink}
\end{equation}
where
\[
S(t)=-\frac{1}{t}\left[\left.\frac{d}{ds}\log h(s)\right\vert_{s=t}\right],
\]
 $h(s)=\int \N(s\vert \tau,1)\pi_{\tau}(\tau)\d\tau$ and 
$\pi_{\tau}(\tau)=
[\SE/\kappa]\,\pi_S([\SE/\kappa]\tau)
$.
\end{prop}

%In the case of the scale mixture of normals considered in this paper, we have $\pi_{\beta}(\beta)=\int \N(\beta\vert 0,\Psi)g(\Psi)\,d\Psi$ and so $\pi_{\tau}(\tau)=\int \N(\tau\vert %0,\Psi^{\star}) g_{\Psi^{\star}}(\Psi^{\star})\,d\Psi^{\star}$ where $g_{\Psi^{\star}}(\Psi^{\star})=
%\SE^2g(\SE^2\Psi)$. Returning to our specific cases, the normal-gamma prior leads to $\Psi^{\star}\sim\Ga(\lambda,\gamma_i/\SE^2)$
%and $\Psi^{\star}\sim\Ga(\lambda,\gamma^{\star}_i);$ with $\gamma^{\star}_i\sim\Ga(c,d\,\SE^2)$ for the normal-gamma-gamma prior.

 Therefore, the shrinkage induced by the posterior expectation (relative to the least squares estimate) can be expressed in terms of a scale defined relative to the standard error. This simplifies the presentation of the shrinkage function for different choices of prior as they can be presented relative to a standard scale. The effect of changing the standard error or the scale of the prior distribution is just to re-scale the $x$-axis of the graphs. The amount of shrinkage depends on various  characteristics of the prior but as a function of prior variance and sampling variance only through the ratio of these as is also the case in simple ridge regression with a normal prior.

\begin{figure}[h!]
\begin{center}
\includegraphics[scale=1, clip, trim = 10mm 0mm 60mm 170mm]{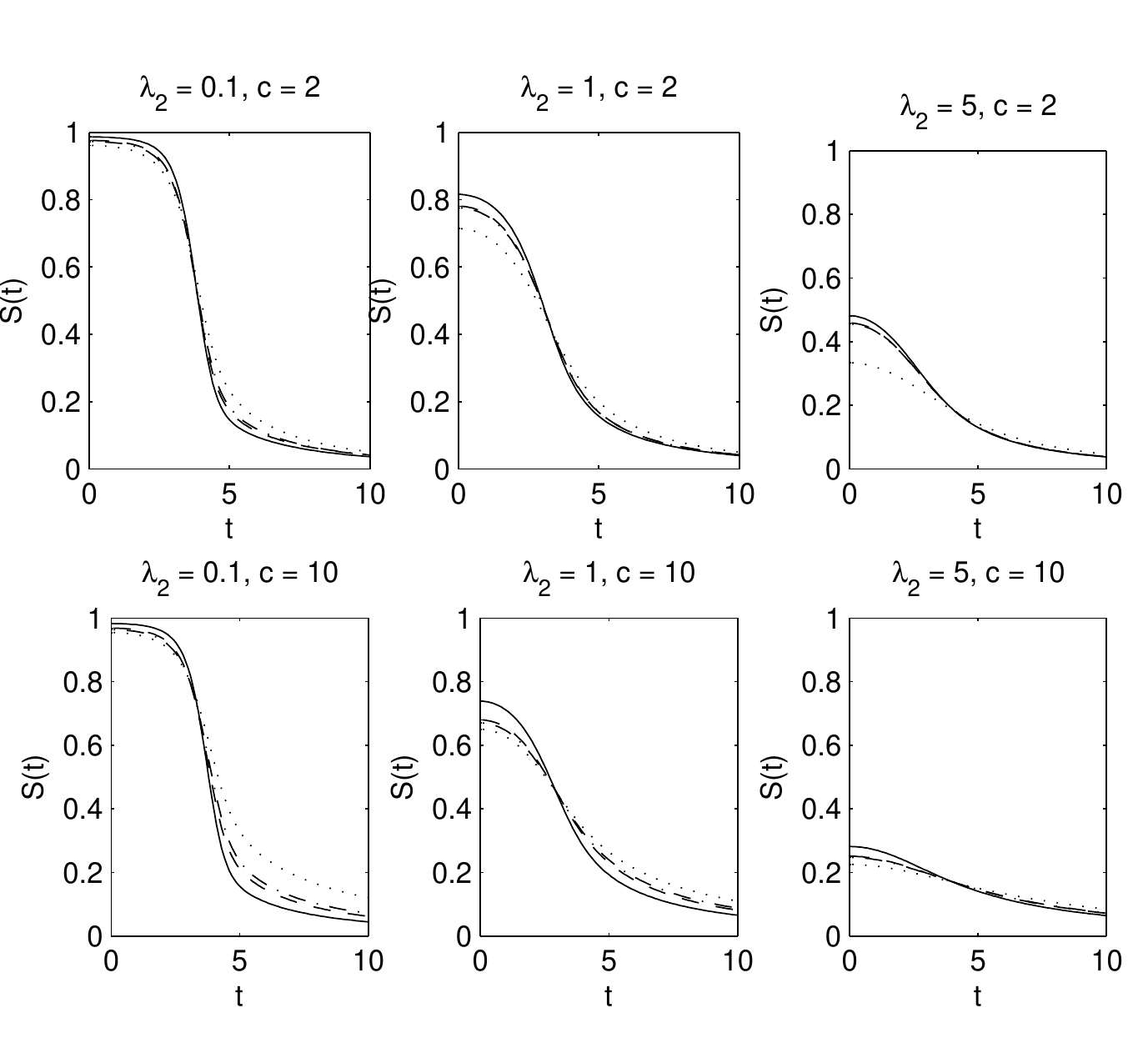} 
\end{center}
\caption{\small
Shrinkage profiles for various choices of products of two normal-gamma-gamma prior distribution with:
 $\lambda_1=\lambda_2$ (solid line), $\lambda_1 = 5\lambda_2$ (dashed line),  $\lambda_1 = 10\lambda_2$ (dot-dashed line) 
 with $d=1/\SE^2$
 compared to a normal-gamma-gamma with  shape $\lambda_1$ (dotted line) with $d=1/\SE^2$.}\label{fig:comp_3}
\end{figure}

%\begin{figure}[h!]
%\begin{center}
%\includegraphics[scale=1, clip, trim = 10mm 0mm 60mm 180mm]{NGG3_prod} 
%\end{center}
%\caption{\small Shrinkage profiles for various choices of product of three normal-gamma-gamma prior distributions where %$\lambda_3=\lambda_2$ with:   $\lambda_2=\lambda_1$ (solid line),
% $\lambda_2 = 5\lambda_1$ (dashed line), 
%$\lambda_2 = 10\lambda_1$
%(dot-dashed line) 
%with $d=1/\SE^2$
%compared to a normal-gamma-gamma with shape $\lambda_1$ (dotted line)
%with $d=1/\SE^2$.}\label{fig:comp_4}
%\end{figure}

Theorems 1 and 2 relate to the shape of the prior density for $\Psi_i$ close to zero when it is defined through products or sums. The appropriateness of the marginal sparsity shape parameter can be checked by comparing 
the shrinkage profiles for a product or a sum of normal-gamma (or normal-gamma-gamma) distributed random variables and for a single normal-gamma (or normal-gamma-gamma) distributed random variable with the marginal sparsity shape parameter of the product or sum.
 If the concept of marginal sparsity is useful then we would expect  the shrinkage profiles to be similar. We consider the following simple prior for coefficients at two levels $\beta^{(1)}$ and $\beta^{(2)}$,
\begin{eqnarray}
\beta^{(1)}\sim\N\left(0, \lambda_1\,d\,\Psi^{(1)}\right),&&\qquad \Psi^{(1)}\sim\Ga(\lambda_1,\lambda_1)
\nonumber\\
\beta^{(2)}\sim\N\left(0,\lambda_2\,d\,\Psi^{(1)}\Psi^{(2)}\right),&& \qquad \Psi^{(2)}\sim\Ga(\lambda_2, \lambda_2) 
\label{model1}
\end{eqnarray}
where $\lambda_2<\lambda_1$. This implies a marginal sparsity shape parameter of $\beta^{(2)}$ is $\lambda_2$.  
The shrinkage for $\beta^{(1)}$ depends on $\lambda_1$ and is unaffected by the choice of $\lambda_2$. However, the shrinkage of $\beta^{(2)}$ depends on both $\lambda_1$ and $\lambda_2$. For comparison, we consider the prior,
\[
\beta^{(2)}\sim\N\left(0, \lambda_2\,d\,\Psi\right), \qquad \Psi\sim\Ga(\lambda_2, \lambda_2)
\]
which has the same prior mean and sparsity shape parameter for $\beta^{(2)}$.

Figure~\ref{fig:comp_1}  shows the shrinkage profiles for different choices of products of two normal-gamma distributions respectively with $d=1/\SE^2$. The marginal sparsity shape parameter is $\lambda_2$ and the shrinkage curve for a single normal-gamma prior with  sparsity shape parameter of $\lambda_2$ is also shown.
Note that here and later as defined in (\ref{shrink}),     $S(t)$  near 1 denotes high shrinkage to zero whereas near zero provides very little shrinkage. Typically we want high shrinkage for small coefficients ($t$ small) and little shrinkage of large coefficients ($t$ large). The shape of the shrinkage curves are very similar for different choices of $\lambda_2$ with shrinkage decreasing slightly as $\lambda_2$ becomes larger. The effect is more pronounced if $\lambda_2$ is smaller. 
%The results with the product of three normal-gamma distributions are similar.
 This suggests that the sparsity shape parameter (although fairly crude)  does give comparable forms of shrinkage for different values of $t$.
Figure \ref{fig:comp_3} show similar graphs for the NGG case with different values of $c$ which show results that are very similar to the normal-gamma case.

Returning to the linear model with interactions, in general we would assume that $\lambda_2<\lambda_1$ since the interactions will tend to be sparser than the main effects. This implies that the {\it marginal} and {\it conditional} sparsity shape parameter of the main effects is $\lambda_1$ and the 
 {\it marginal} and {\it conditional} sparsity shape parameter of the interactions is $\lambda_2$.
In the GAM model,  the marginal sparsity shape parameters of the basis functions for the $j$-th variable are $\min\{\lambda_1,\lambda_{2,j}\}$, and the conditional sparsity shape parameter of the basis functions for the $j$-th variable is $\lambda_{2,j}$. The marginal and conditional sparsity shape parameter of the variables is $\lambda_1$.

\subsection{Shape versus scale shrinkage} 

We have emphasised the importance of the sparsity shape parameter to achieve an effective control of sparsity induced by the prior. An alternative specification would induce extra shrinkage at higher levels through a prior with the {\it same} sparsity shape parameter but {\it different} scales at different levels (for example, directly extending the approach of \cite{BhDu11} to this situation). Such a model could be expressed as
\begin{eqnarray}
\beta^{(1)}\sim\N(0, \lambda_1\,d\,\Psi^{(1)}),&&\qquad \Psi^{(1)}\sim\Ga(\lambda_1,\lambda_1)
\nonumber\\
\beta^{(2)}\sim\N\left(0,\lambda_2\,d\,\Psi^{(1)}\Psi^{(2)}\right),&& \qquad \Psi^{(2)}\sim\Ga(\lambda_1, \lambda_1).\label{model2}
\end{eqnarray}
The coefficients $\beta^{(1)}$ and $\beta^{(2)}$ have the same prior variance as in the model in (\ref{model1}). We refer to this prior, (\ref{model2}), as Scale-Induced Shrinkage (ScIS) and our model in  (\ref{model1}) as Shape-Induced Shrinkage (ShIS).

\begin{figure}[h!]
\begin{center}
\includegraphics[scale=1, clip, trim = 10mm 0mm 60mm 230mm]{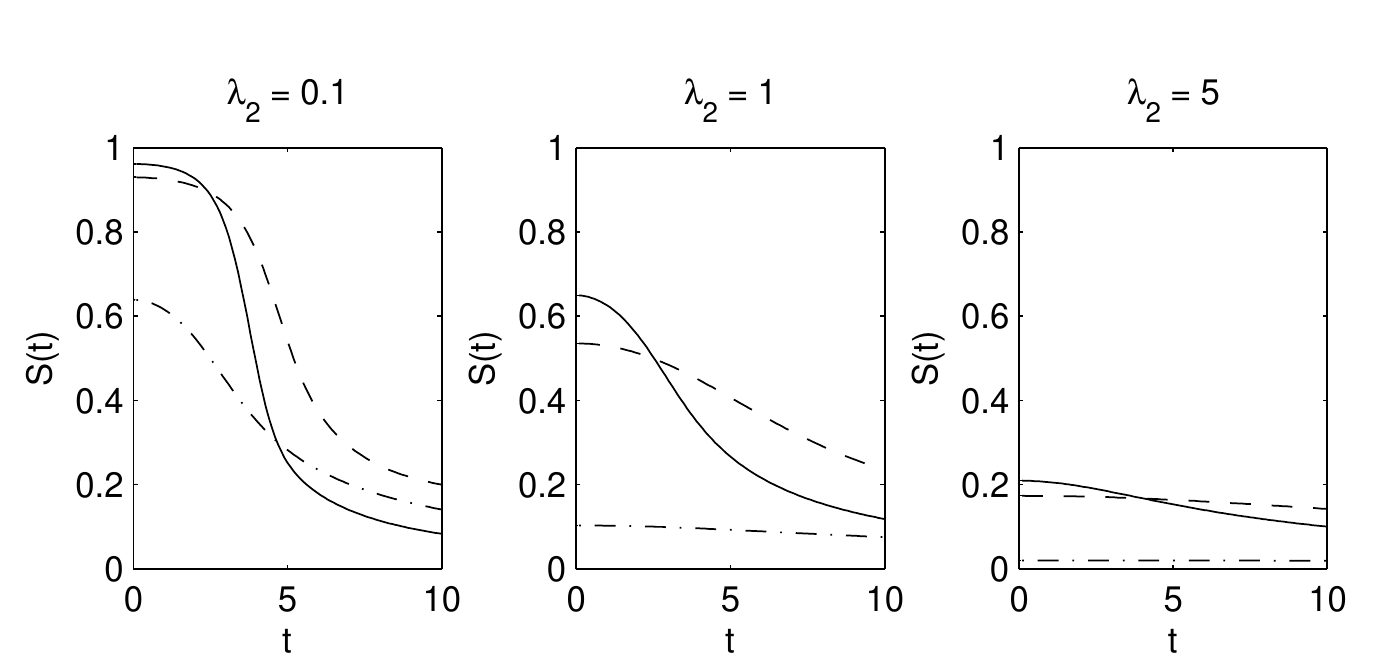} 
\end{center}
\caption{\small
Shrinkage profiles for the model with the ShIS and ScIS priors  with $\lambda_1=10\lambda_2$. The shrinkage profile are for: $\beta^{(2)}$ with the ShIS prior (solid line),
$\beta^{(2)}$ with the ScIS   prior (dashed line),  and $\beta^{(1)}$ with either prior (dot-dashed line) 
 with $d=1/\SE^2$.}\label{fig:comp_4}
\end{figure}
Figure~\ref{fig:comp_4} shows the shrinkage profile for both the ShIS and ScIS priors. The ShIS prior  leads to more adaptive shrinkage than the ScIS  prior, that is  more shrinkage for small coefficients and and less shrinkage  for larger coefficients,  and the effect is more pronounced when $\lambda_2$ is small. Smaller $\lambda_2$ indicates greater sparsity which is the types of priors in which we are particularly interested. The shape of the shrinkage profiles for $\beta^{(2)}$ with the ScIS prior more closely resemble the shrinkage profiles for $\beta^{(1)}$ but with a re-scaling due to the smaller prior mean.

\section{Computational strategy}
\label{se:comp}

Posterior inference with these priors can be made using Markov chain Monte Carlo methods. In this section, we will describe the general strategy for inference rather than describe algorithms for specific models. We will assume the general model 
\[
y_i=\alpha+\sum_{l=1}^L X^{(l)}_{i}\beta^{(l)} + \epsilon_i,\qquad i=1,\dots,n
\] 
where $X^{(l)}_i$ is a $(n\times p_l)$-dimensional matrix whose columns are given by the variables in the $l$ level and $\epsilon_i
\stackrel{i.i.d.}{\sim}\N(0,\sigma^2)$,
\[                                                  
\beta^{(l)}_j\stackrel{i.i.d.}{\sim}\N\left(0,\Psi^{(l)}_j\right),\qquad 
j=1,\dots,p_l, \quad l=1,\dots,L
\]
and
\begin{equation}
\Psi^{(l)}_j = s_j^{(l)}\,d\,\frac{f_{jl}\left(\Psi^{(1)},\dots,\Psi^{(l-1)}\right)}{\E[f_{jl}\left(\Psi^{(1)},\dots,\Psi^{(l-1)}\right)]}
\,\eta^{(l)}_j,\qquad 
j=1,\dots,p_l,\quad l=1,\dots,L.
\end{equation}
Typically,  the distribution of $\eta_j^{(l)}$ has parameters which are denoted  $\phi^{(l)}$. The Gibbs sampler will be used to sample from the posterior distribution of the parameters 
$(\alpha,\beta,\sigma,\Psi,d,\phi)$ where $\beta=\{\beta^{(l)}\vert l=1,\dots,L\}$, $\Psi=\{\Psi^{(l)}\vert l=1,\dots,L\}$ and 
$\phi=\{\phi^{(l)}\vert l=1,\dots,L\}$. The full conditional distributions of 
$(\alpha,\beta)$ and $\sigma^2$ follow from standard results for Bayesian linear regression models. The parameters $\Psi$, $d$ and $\phi$ are updated  one-element-at-a-time by adaptive Metropolis-Hastings random walk steps using a variation on the algorithm proposed by \cite{atros05}. The output of adaptive Metropolis-Hastings algorithms are not Markovian (since the proposal distribution is allowed to depend on the previous values of the Markov chain) and so standard Markov chain theory cannot be used to show that the resulting chain is ergodic. Relatively simple conditions are given for the ergodicity of adaptive Metropolis-Hastings algorithms by \cite{RoRo07}. Our algorithms meet these conditions with the additional restriction that $\Psi$, $d$ and $\phi$ are bounded above (at a very large value). Suppose that we wish to update $\phi^{(l)}$ at iteration $i$ (the same idea will also be used to update the elements of $\Psi$ and $d$). A new value $\phi^{(l)\ '}$ is proposed according to
\[
\log\phi^{(l)\ '}=\log\phi^{(l)} + \epsilon^{(l)}
\]
where $\epsilon^{(l)} \sim \N\left(0, \sigma^{2\ (i)}_{\phi^{(l)}}\right)$. The notation $\sigma^{2\ (i)}_{\phi^{(l)}}$ makes the dependence on the previous 
values of the chain explicit and the induced transition density of the  proposal  is denoted
$q_{\sigma^{2\ (i)}_{\phi^{(l)}}}\left(\phi^{(l)},\phi^{(l)\ '}\right)$.
The value $\phi^{(l)\ '}$ is accepted or rejected using the standard Metropolis-Hastings acceptance probability
\[
\alpha\left(\phi^{(l)},\phi^{(l)\ '}\right)
=\frac{\prod_{j=1}^{p_l} p\left(\left.\Psi^{(l)}_j\right\vert \phi^{(l)\ '}\right)
p\left(\phi^{(l)}\ '\right)\,q_{\sigma^{2\ (i)}_{\phi^{(l)}}}\left(\phi^{(l)\ '},\phi^{(l)}\right)}{\prod_{j=1}^{p_l} p\left(\left.\Psi^{(l)}_j\right\vert \phi^{(l)}\right)p\left(\phi^{(l)}\right)\,q_{\sigma^{2\ (i)}_{\phi^{(l)}}}\left(\phi^{(l)},\phi^{(l)\ '}\right)}.
\]
The variance of the increment is updated by
\[
\log\sigma^{2\,(i+1)}_{\phi^{(l)}}=
\log\sigma^{2\,(i)}_{\phi^{(l)}}+i^{-a}\left(\alpha\left(\phi^{(l)},\phi^{(l)\ '}\right)-\tau\right)
\]
where $1/2<a\leq 1$. This algorithm leads to an average acceptance rate  which converges to $\tau$. 
We choose $a=0.55$ and $\tau=0.3$ (following the suggestion of \cite{RoRo09}) in our examples.

The posterior distribution can be highly multi-modal and so it is necessary to use parallel tempering to improve the mixing. 
An effective, adaptive implementation is described by
\cite{MiMoVi12}.

\section{Examples}
\label{se:example}

\subsection{Example 1: Prostate cancer data}

Data from a prostate cancer trial \citep{Stamey} have become a standard example in the regularization literature
\citep{tib96, zouhastie05, Kyung10}. The response is the logarithm of prostate-specific antigen ({\it lpsa}).
There are eight predictors: log(cancer volume) ({\it lv}), 
log(prostate weight) ({\it lw}), age (in years), the logarithm of the amount of benign prostatic hyperplasia ({\it lbph}, log(capsular penetration) ({\it lcp}), Gleason score ({\it gl}),
percentage Gleason score 4 or 5 ({\it pg})), and
seminal vesicle invasion ({\it svi}). 

We considered all  variables to be continuous apart from {\it svi} which is binary (it should be noted that Gleason score is ordinal and has 4 observed  levels (scores of 6, 7, 8 and 9) in the data).
 Previous modelling had often included the continuous variables as linear effects. An exception is \cite{LaiHuaLee12} who considered flexibly modelling their effects. We followed this approach using the GAM model in section~\ref{sse:GAM} with the prior described in 
section~\ref{se:GAM_prior}. All continuous variables were normalized to have a minimum of 0 and a maximum of 1. A piecewise linear spline basis function was assumed for the $j$th variable, $j=1,\ldots, p,$ so that
\[
f_j(x_{ij}) = \theta_j x_{ij} + \sum_{k=1}^K[ (x_{ij}-\tau_k)_+ \gamma_{jk}]
\]
where $(x)_+=\max\{0,x\}$ and $\tau_k=\frac{k-1}{K-1}$ for $k=1,\dots,K$. In this example, we use $K=60$. The priors for the hyperparameters were:  $\lambda_1\sim\Ga(1,1)$, $\lambda_{2,j}\stackrel{i.i.d.}{\sim}\Ga(1,10)$, 
 and $p(d)\propto (1+d)^{-2}$. The parameter $\lambda_1$ controls sparsity at the variable level and  the choice centres the prior for the regression coefficients over the Bayesian lasso prior. The smaller prior mean for $\lambda_{2,j}$, $\E[\lambda_{2,j}]=0.1,$ implies greater sparsity at the basis level than the variable level and that only a few knots will be important for each variable.

\begin{figure}[h!]
\begin{center}
 \includegraphics[trim=10mm 0mm 70mm 190mm, scale=1, clip]{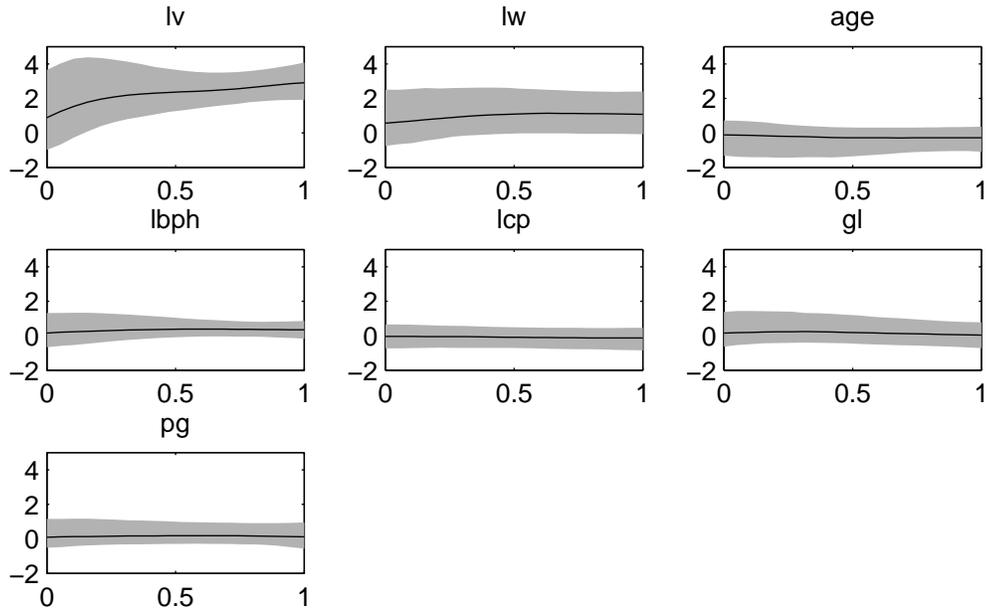} 
\end{center}
\caption{\small Prostate cancer data -- the posterior distribution of the linear effects $\beta_j(x)$ for each variable summarized as the posterior median (solid line) and pointwise $95\%$ credible interval (grey shading)}
\label{f:results_prostate}
\end{figure}
The results of fitting the flexible regression model are shown in Figure~\ref{f:results_prostate}. 
The inference about the regression effects are shown as $\beta_j(x)=\frac{f_j(x)}{x}$ and can be interpreted as the variable-dependent linear regression effect for the $j$-th variable.  The effect of {\it lv} was clearly important with an effect with the posterior median increasing from 0.88 to 2.91 over the range of the data. The effect of {\it lw} also seemed important and relatively constant over the range of the data. The other variables were clearly less important with a posterior median which is constant and close to zero and a narrower $95\%$ credible intervals than the other variables. The effect of {\it svi} had a posterior median of 0.58 with a 95\% credible interval of $(0.08, 1.06)$ which indicated the importance of this variable for the regression model.

\begin{figure}[h!]
\begin{center}
\includegraphics[trim=10mm 0mm 70mm 220mm, scale=0.8, clip]{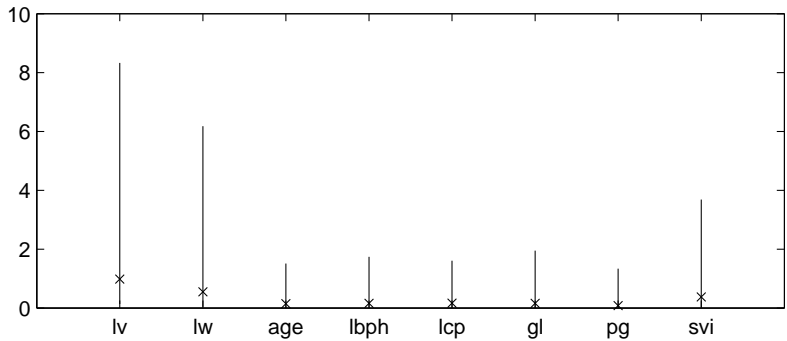}
\end{center}
\caption{\small Prostate cancer data -- the posterior distribution of $\Psi$ for each variable
summarized as the posterior median (cross) and $95\%$ credible interval (solid line)}
\label{Psi_prostate}
\end{figure}
The posterior distribution of  the $\Psi^{(1)}_i$ is a measure of the overall strength of effect for the $i$-th variable. The distribution for each variable is shown in Figure~\ref{Psi_prostate}. The results were consistent with the estimates of the regression effects. The {\it lv} variable gave the largest posterior median and had support at larger values of $\Psi$ than other variables. 
The variables {\it lw} and {\it svi} also had important effects and had the next two largest values of the posterior median and were clearly useful as a scalar summary of the regression effects.

\begin{table}[h!]
\begin{center}
\begin{tabular}{cc}\hline
$\lambda_1$ & 0.96 {\it (0.31, 3.44)}\\
$d$ & 0.64 {\it (0.09, 6.10)} \\\hline
\end{tabular}
\end{center}
\caption{\small Prostate cancer data -- the posterior distribution of the hyperparameters summarised as posterior median and 95\% credible interval}\label{t:prostate_inf}
\end{table}

\begin{figure}[h!]
\begin{center}
\includegraphics[trim=10mm 0mm 70mm 220mm, scale=0.8, clip]{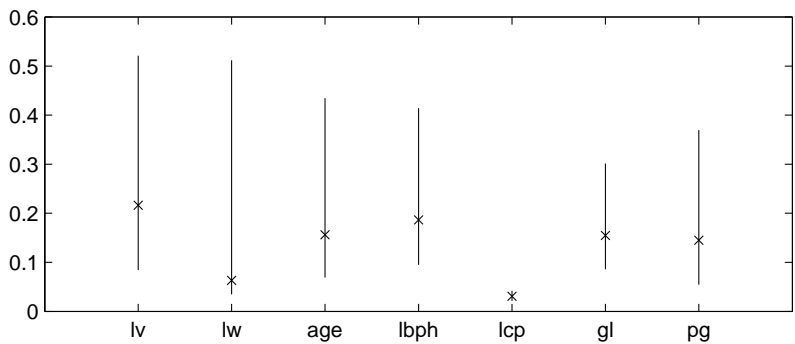}
\end{center}
\caption{\small Prostate cancer data -- the posterior distribution of $\lambda_{2,1},\dots,\lambda_{2,7}$ 
summarized as the posterior median (cross) and $95\%$ credible interval (solid line)}
\label{lambda_prostate}
\end{figure}
A summary of the posterior distribution of $\lambda_1$ and $d$ are shown in Table~\ref{t:prostate_inf} and a summary of the posterior distributions of variable-specific basis  level sparsity parameters, $\lambda_{2,j}$, are shown in Figure~\ref{lambda_prostate}. The posterior median of $\lambda_1$ is close to 1 indicating that only some of the variables are important but that there is not a high degree of sparsity. The parameter $\lambda_{2,j}$ indicates the sparsity in the coefficients of the spline  basis for the $j$-th variable. A smaller value of $\lambda_{2,j}$ indicates fewer splines are needed to model the effect of the variable and, therefore, are a measure of the departure from linearity for each variable. The variable {\it lv} has the largest posterior median and so the largest departure from linearity whereas {\it lcp} has the smallest posterior median and so the smallest departure from linearity. This is consistent with the estimated effects shown in 
Figure~\ref{f:results_prostate}.

\subsection{Example 2: Computer data}

Data on the characteristics and performance of 209 CPUs were considered by \cite{einfel87} and subsequently analysed by \cite{gus00} using Bayesian non-linear regression techniques. The response is performance of the CPU. In common with \cite{gus00}, we consider 5 predictors: A, the machine cycle time (in nanoseconds); B, the average main memory size (in kilobytes); C, the cache memory size (in kilobytes); D, the minimum number of input channels; and E, the maximum number of input channels. In a similar spirit to \cite{gus00}, we modelled the data using a GAM with interactions which introduces bivariate functions, $f_{jl}(\cdot,\cdot)$, which allows modelling of non-linear interaction effects. In this case, 
 the GAM model is extended to
\begin{align}
y_i=\, &\sum_{j=1}^p f_j(X_{ij})+\sum_{j=1}^p \sum_{k=1}^{j-1} f_{jk}(X_{ij},X_{ik})+
\epsilon_i\nonumber\\
=\,&
 \sum_{j=1}^p \theta_{j}^{(M)} X_{ij}+
\sum_{j=1}^p \sum_{k=1}^{K} \gamma_{jk}^{(M)} g(X_{ij},\tau_{jk})+
 \sum_{j=1}^p\sum_{k=1}^{j-1}
 \theta_{jk}^{(I)} X_{ij} X_{ik}\nonumber\\
&+
\sum_{j=1}^p \sum_{k=1}^{j-1}
\sum_{l=1}^{K}\sum_{m=1}^{K} \gamma^{(I)}_{jklm} 
 g(X_{ij},\tau_{jl}) g(X_{ik},\tau_{km})+
\epsilon_i
\label{GAM_interaction}
\end{align}
where, again, $\epsilon_i\sim\N(0,\sigma^2)$.
The $\gamma$ parameters for the nonlinear functions (splines) involve $K$ knots. The bracketed superfixes (M) and (I) refer to main effects and interaction levels respectively.
We used the model with
$g_j(x,\tau) =  (x-\tau)_+$ and $K=10$ knots. 
 The 5 main effects and 10 interactions lead to 1055 regression parameters in the model.

\cite{gus00} 
used a square root transformation of the predictors since these data are highly skewed. In principle the distribution of variables shouldn't matter in non-linear regression modelling. However, knots are evenly spaced and so it would be useful to have data relatively evenly spread across the range of the knots. We found that a log transformation of the response lead to better behaved residuals than the untransformed response and also transformed the variables by $f(x)=\log(1+x)$. All transformed variables were subsequently transformed to have a minimum of 0 and a maximum of 1.

  A hierarchical sparsity prior can be constructed for this problem by combining the prior for a GAM with only main effects and the prior for the linear model with interactions. The regression coefficients are organized into four levels:
a main effects level, an interactions level, a basis level for main effects, and a basis level for interaction.  
 The main effects level has 
  $p_1=p$ terms of the form
$\theta^{(M)}_{j}$ for $j=1,\dots,p$.
The interaction level has $p_2=p(p-1)/2$ terms of the form
$\theta^{(I)}_{jk}$ for $j=1,\dots,p$ and $k=1,\dots,j-1$.
 The basis level for main effects contains 
$\gamma_{jk}^{(M)}$ for $j=1,\dots,p$, $k=1,\dots,K$ and has $p_3=pK$ terms. 
The basis level for interactions contains 
 $\gamma^{(I)}_{jklm}$ for $j=1,\dots,p$, $k=1,\dots,j-1$, $l=1,\dots,K$, $m=1,\dots,K$ and contains 
 $p_4=(p-1)/2 K^2$.
The proposed prior, with strong heredity, is 
\[
\theta^{(M)}_{j}\sim\N\left(0, \lambda_1\,d\,\eta^{(1)}_j\right),\quad
\eta^{(1)}_{j}\sim\GG\left(\lambda_1,c,\frac{c-1}{\lambda_1}\right),
\]
\[
\theta^{(I)}_{jk}\sim\N\left(0, \lambda_2\,d\,\eta^{(2)}_{jk}\,\eta^{(1)}_j\,\eta^{(1)}_k\right),\quad \eta^{(2)}_{jk}\sim\GG\left(\lambda_2,c,\frac{c-1}{\lambda_2}\right). 
\]
\[
\gamma^{(M)}_{jk}\sim\N\left(0, \lambda_3\,d\,\eta^{(3)}_{jk}\,\eta_j^{(1)}\right),\quad
\eta^{(3)}_{jk}\sim\GG\left(\lambda_{3,j},c,\frac{c-1}{\lambda_3}\right).
\]
\[
\gamma^{(I)}_{jklm}\sim\N\left(0,\lambda_{4,j,k}\,d\,
\,\eta^{(4)}_{jklm}
\,\eta^{(2)}_{jk}
\eta_j^{(1)}\eta^{(1)}_k\right),\qquad
\eta^{(4)}_{jk}\sim\GG\left(\lambda_4,c,\frac{c-1}{\lambda_4}\right),
\]

If $\eta_j^{(1)}$ is small then both the main effects $\theta_{j}^{(M)}$ and the basis function coefficients $\gamma_{jk}^{(M)}$ will tend to be small. Similarly, if $\eta_{jk}^{(2)}\eta_j^{(1)}\eta_k^{(1)}$ is small then both the interaction terms $\theta_{jk}^{(I)}$ and the basis function coefficients $\gamma^{(I)}_{jklm}$ will tend to be small. This allows variable selection at the main effect and interaction term levels. The prior also links  the interaction and main effects terms (and, consequently, their associated basis function coefficients) since $\eta_{jk}^{(2)}\eta_j^{(1)}\eta_k^{(1)}$ is more likely to be small if both $\eta_j^{(1)}$ and $\eta_k^{(1)}$ are small. We assume that $\lambda_2<\lambda_1$ and so the  marginal sparsities are 
$\lambda_1$ for the main effects level, $\lambda_2$ for the interactions level, $\min\{\lambda_1,\lambda_{3,j}\}$ for the basis level for the $j$-th main effects and $\min\{\lambda_2,\lambda_{4,j,k}\}$ for the basis level for interactions.

The priors for the hyperparameter of the model were as follows. The sparsity parameters for the main effects and interaction terms were chosen as $\lambda_1\sim\Ex(1)$ and  $\lambda_2=r\lambda_1$ where 
 $r\sim\Be(2,6)$ which implied that $\E[r]=1/3$ suggesting that the interaction are {\it a priori} much sparser than the main effects. The conditional sparsity shape parameters for the nonlinear terms were chosen to be $\lambda_{4,j,k}\stackrel{i.i.d.}{\sim}\Ga(1,100)$ and $\lambda_{2,j}\stackrel{i.i.d.}{\sim} \Ga(1,10)$ which implies that nonlinear terms were less likely to be included in the interaction function than the main effects function (which reflected the larger number of terms in the interaction function).
  The scale parameter, $d,$ was given the prior $p(d)\propto (1+d)^{-2}$ which implied that $\E[d]=1$ but with a heavy tail.

\begin{figure}[h!]
\begin{center}
\begin{tabular}{c}
Main effects\\
\includegraphics[trim=10mm 0mm 55mm 235mm, scale=0.9, clip]{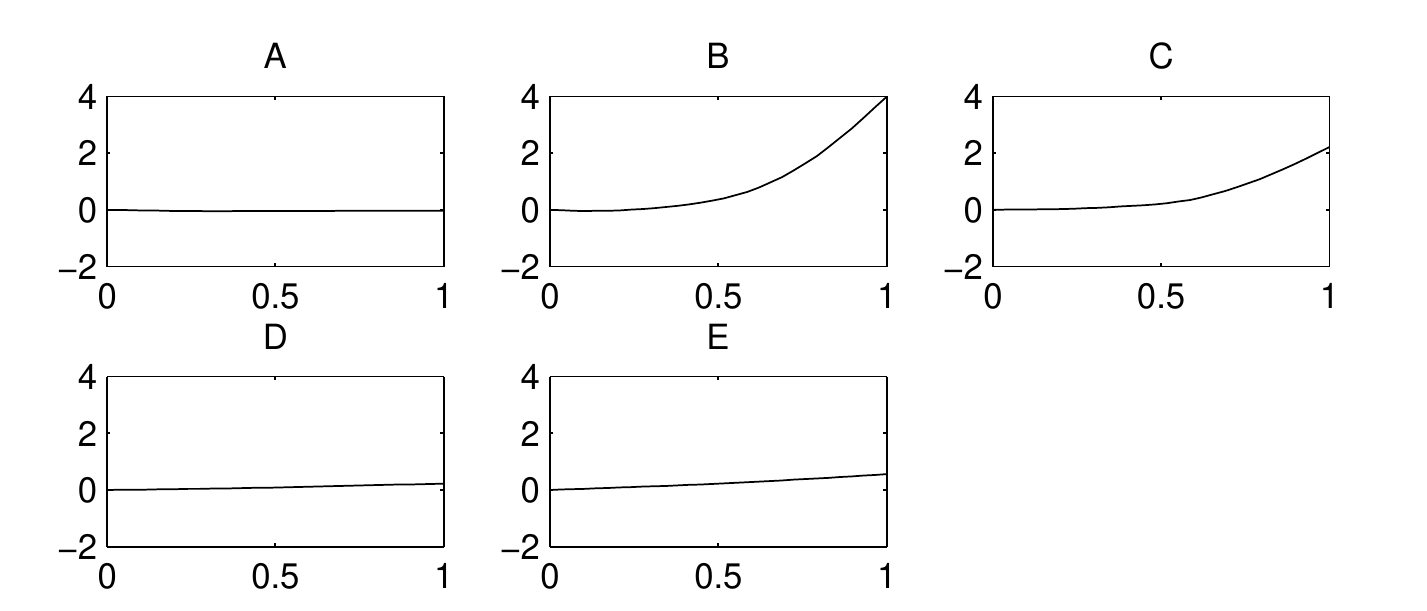}\\
\\
Interactions\\
\includegraphics[trim=10mm 0mm 55mm 180mm, scale=0.9, clip]{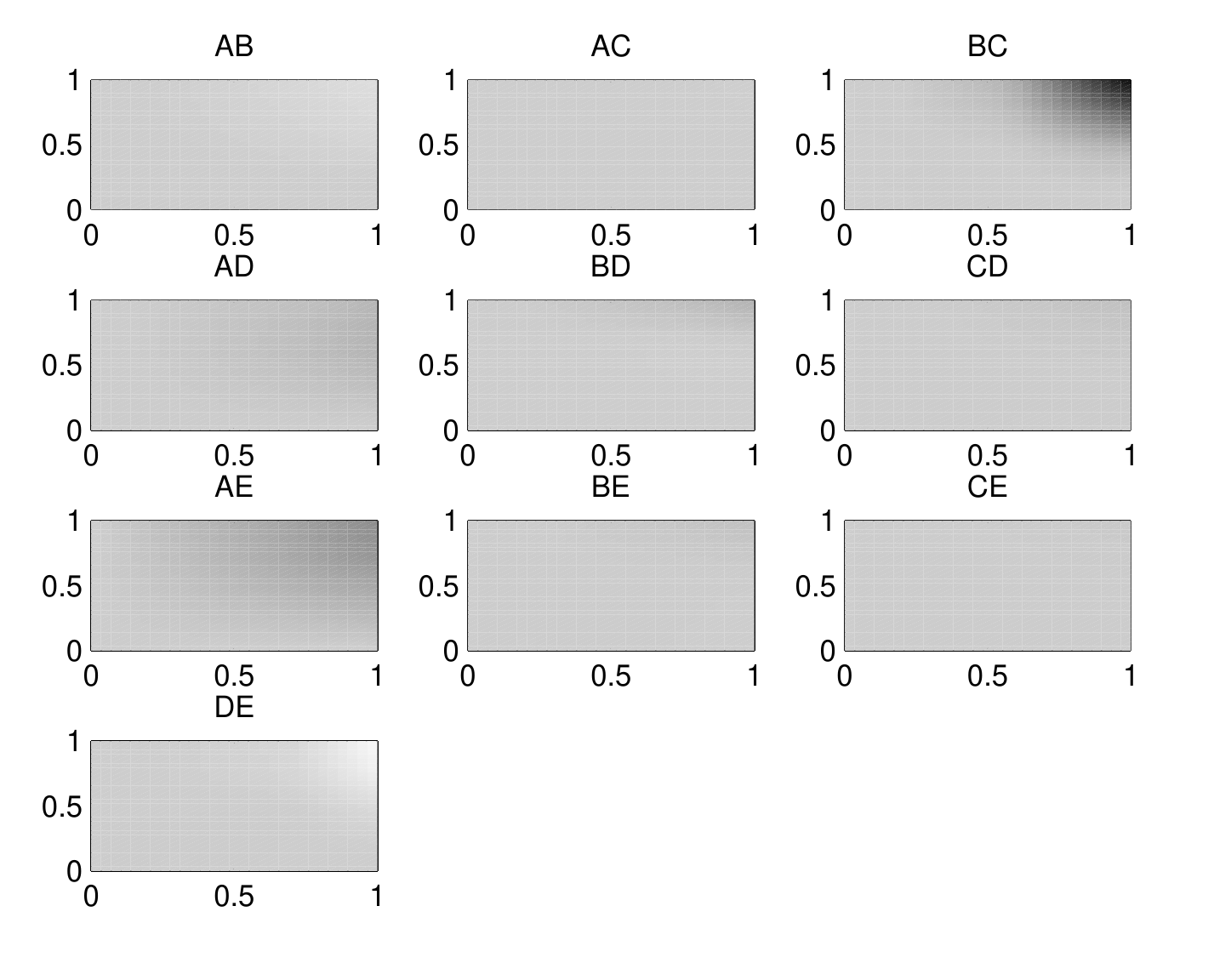}
\end{tabular}
\end{center}
\caption{\small Computer data -- the posterior mean of each main effect and each interaction. Darker colours represent lower values for the interaction graphs.}
\label{beta_cpu}
\end{figure}

The estimated main effects and interactions are shown in Figure~\ref{beta_cpu}. The effect of A, D and E were small whereas B and C had an increasing, non-linear effect with a largest effect of roughly 4 for B and roughly 2 for C. The interaction effects mostly had a posterior median of zero. The main exception was the interaction between B and C which has a posterior median of -4 when both B and C are 1. This indicated that the effect of large values of B and C were over-estimated by the linear effects alone.

\begin{figure}[h!]
\begin{center}
\begin{tabular}{cc}
Main effects & Interactions\\
\includegraphics[trim=0mm 0mm 135mm 240mm, scale=1, clip]{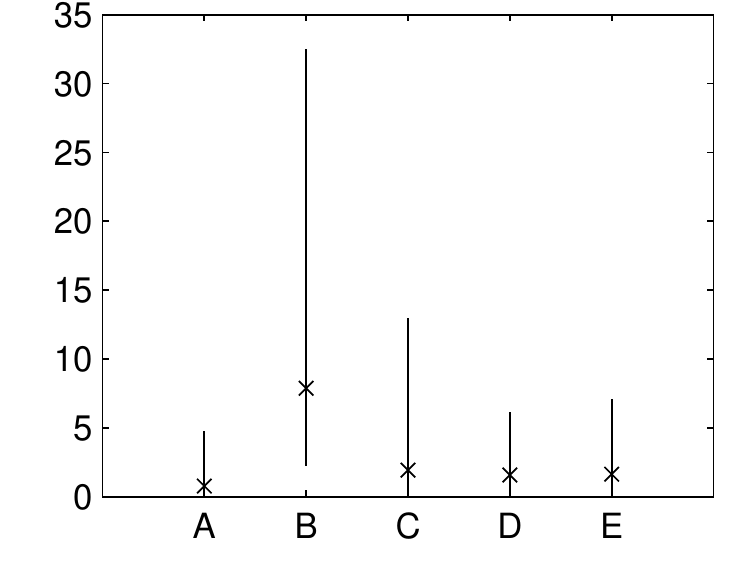} &
\includegraphics[trim=0mm 0mm 135mm 240mm, scale=1, clip]{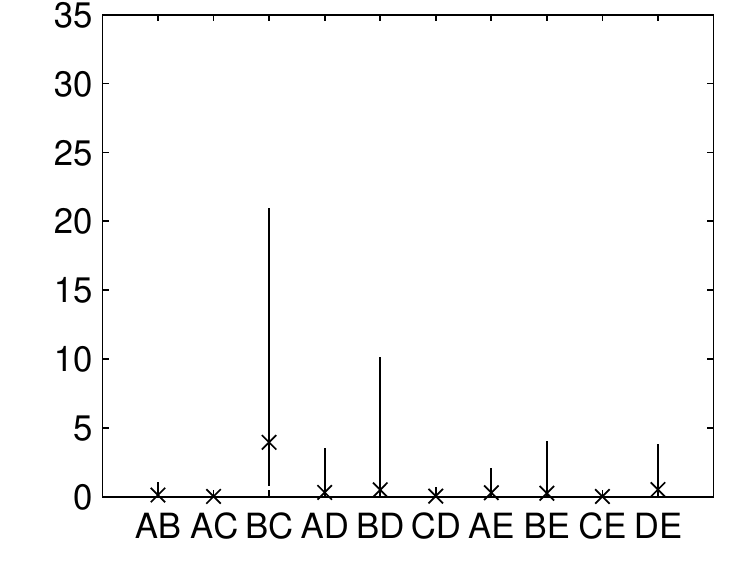}
\end{tabular}
\end{center}
\caption{\small Computer data -- the posterior distribution of $\Psi$ for each main effect and each interaction summarized as the posterior median (cross) and $95\%$ credible interval (solid line).}
\label{Psi_cpu}
\end{figure}
Figure~\ref{Psi_cpu} shows the posteriors for the $\Psi$'s for the main effects and interactions. 
These results were consistent with the estimated effects. The variables B and C had the largest posterior medians and upper point of the 95\% credible interval
for the main effects. Similarly, the interaction between B and C had the largest posterior median and upper point of the 95\% credible interval than the other interactions. 

\begin{table}[h!]
\begin{center}
\begin{tabular}{cc}\hline
$\lambda_1$ & 1.96 {\it (0.41, 4.68)}\\
$\lambda_2$ & 0.40 {\it (0.13, 1.12)}\\
$d$ & 0.84 {\it (0.09, 10.06)}\\\hline
\end{tabular}
\end{center}
\caption{\small Computer data -- the posterior distribution of the hyperparameters
summarised as posterior median and 95\% credible interval}
\label{t:cpu_inf}
\end{table}

%\begin{figure}[h!]
%\begin{center}
%\begin{tabular}{cc}
%$\lambda_3$ & $\lambda_4$\\
%\includegraphics[trim=0mm 0mm 135mm 240mm, scale=1, clip]{cpus_lambda2}  &
%\includegraphics[trim=0mm 0mm 135mm 240mm, scale=1, clip]{cpus_lambda2_interactions} 
%\end{tabular}
%\end{center}
%\caption{\small Computer data -- the posterior distribution of $\lambda_3$ and $\lambda_4$ for each main effect and each %interaction respectively summarized as the posterior median (cross) and $95\%$ credible interval (solid line) using four %different priors.}
%\label{lambda_cpu}
%\end{figure}
A summary of the posterior distribution of $\lambda_1$, $\lambda_2$ and $d$ are shown in Table~\ref{t:cpu_inf}.
% and a summary of the posterior distributions of variable-specific $\lambda_{4,j,k}$ and $\lambda_{2,j}$ are shown in Figure~\ref{lambda_cpu}. 
The posterior median of $\lambda_1$ is close to 2 which indicates that most effects are relatively important (although this is estimated with a wide 95\% credible interval due to the small number of regressors). The posterior median of $\lambda_2$ indicates that the interactions are much sparser than the main effects. 
%The sparsity parameters $\lambda_{3,j}$ and $\lambda_{4,j,k}$ indicate the amount by which the effects deviate from linearity. The posterior medians of $\lambda_{2,j}$ for B and C are much larger than for the other variables. This indicates a departure from linearity which is confirmed by the estimate regression effects in Figure~\ref{beta_cpu}. The posterior distributions of $\lambda_{4,j,k}$ are fairly similar indicating little difference in the level of departure from normality for the interactions (the relatively small effect of the interactions leads to a small amount of information about these parameters).

\subsection{Out-of-sample predictive performance}

The performance of the hierarchical prior introduced in this paper was compared using  five-fold cross-validation to three priors which do not assume dependence between the regression coefficients.
\begin{table}[h!]
\begin{center}
\begin{tabular}{ccccc}\hline
 & \multicolumn{2}{c}{Prostate cancer}
 & \multicolumn{2}{c}{Computer data}\\
&                   RMSE    &    LPS &        RMSE    &    LPS\\\hline
Hierarchical  & {\bf 0.7946}  &     {\bf 1.1830} &   {\bf 0.037 }  &  {\bf -2.55}\\
NGG         &  0.8237     &  1.2154 &      2.119  &    0.70\\
HS           &  1.1083       & 1.6496     &   1.204 &     1.79\\
SSVS         & 0.8518      &  1.2394       & 0.043   &  -1.91 \\\hline
\end{tabular}
\end{center}
\caption{The root mean squared errors (RMSE) and log predictive scores (LPS) wih the prostate cancer and computer data examples. The smallest value of each measure is shown in bold.}
\end{table}
These were: a ``spike-and-slab'' prior, normal-gamma-gamma prior and horseshoe prior. The results are summarized by both the root mean squared error (RMSE) where the posterior predictive median was used as the estimated prediction and the log predictive score \citep{good52}. The posterior predictive median (rather than mean) was used since the heavy-tailed priors tended to produce heavy-tailed predictive distribution which were better summarized by the median. The hierarchical prior has a smaller RMSE and LPS than the priors with no dependence for both data sets.

\section{Discussion} \label{se:disc}

This paper describes a hierarchical approach to prior construction in sparse regression problems.  We assume that variables can be divided into levels and the relationship between the regression coefficients can be expressed hierarchically. The framework allows control of both the conditional sparsity and marginal sparsity of groups of regression coefficients at different levels of the prior. Complexity is controlled by manipulating sparsity in the hierarchical prior through notions of strong and weak heredity. This is done through the shape rather than the scale of the gamma-gamma mixing density and as a result gives good adaptivity. These priors have natural applications in problems  such as models with interactions and non-linear Bayesian regression models. These priors are able to find sparse estimates in situations where there are large numbers of parameters. We feel that these approaches will have the potential for many applications in future. For example, \cite{kalgri12} use a simple, two stage  hierarchical prior in a regression model with time-varying regression coefficients. This allows the control of both sparsity of the variables (where  values of the regression coefficients at all times are shrunk to zero) and sparsity of each regression coefficient over time.

\bibliographystyle{chicago}
\bibliography{ref_home}

\appendix

\section{Proofs}

\subsection{Proof of Theorem 1}

\subsubsection*{Part (i)}

Suppose that $\lambda_1= \min\{\lambda_i\}$ then
\[
p(\Psi)=\prod_{i=1}^K \frac{1}{\Gamma(\lambda_i)}
\Psi^{\lambda_1-1}\int_0^{\infty} \cdots \int_0^{\infty}
\exp\left\{-\Psi\left/\prod_{i=2}^K \eta_i\right.\right\}\prod_{i=2}^K
\eta_i^{\lambda_i-\lambda_1-1}\exp\left\{-\sum_{i=2}^K \eta_i\right\}
d\eta_2\cdots d\eta_K
\]
Thus 
\begin{align*}
C(\Psi)&=
p(\Psi)/\Psi^{\lambda_1-1}\\
&=\prod_{i=1}^K \frac{1}{\Gamma(\lambda_i)}
\int_0^{\infty} \cdots \int_0^{\infty}
\exp\left\{-\Psi\left/\prod_{i=2}^K \eta_i\right.\right\}\prod_{i=2}^K
\eta_i^{\lambda_i-\lambda_1-1}\exp\left\{-\sum_{i=2}^K \eta_i\right\}
d\eta_2\cdots d\eta_K
\end{align*}

By the dominated convergence theorem

\begin{align*}
\lim_{\Psi\rightarrow 0} C(\Psi)= C(0)=
\prod_{i=1}^K \frac{1}{\Gamma(\lambda_i)}
\prod_{i=2}^K
\int_0^{\infty}
\eta_i^{\lambda_i-\lambda_1-1}\exp\left\{-\sum_{i=2}^K \eta_i\right\}
d\eta_i
=\frac{1}{\Gamma(\lambda_1)}
\prod_{i=2}^K \frac{\Gamma(\lambda_i-\lambda_1)}{\Gamma(\lambda_i)}
\end{align*}
since $\lambda_i\geq \lambda_1$. Therefore, the sparsity shape parameter is $\min\{\lambda_i\}$.
\hfill{$\square$}

\subsubsection*{Part (ii)}

In this case, $\Psi\sim\Ga(\sum_{i=1}^K \lambda_i, 1)$ and so the sparsity shape parameter is $\sum_{i=1}^K \lambda_i$.

\subsection{Proof of Theorem 2}

\subsubsection*{Part (i)}

Suppose that $\lambda_1= \min\{\lambda_i\}$ then
\[
p(\Psi)%=\left\{\prod_{i=1}^K \frac{1}{\mbox{Beta}(\lambda_i,c_i)}\right\}
\propto
\Psi^{\lambda_1-1}\int_0^{\infty} \cdots \int_0^{\infty}
\left\{1+\Psi\left/\prod_{i=2}^K \eta_i\right.\right\}^{-(\lambda_1+c_1)}\left\{\prod_{i=2}^K
\eta_i^{\lambda_i-\lambda_1-1}\left\{1+\eta_i\right\}^{-(\lambda_i+c_i)}\right\}d\eta_2\cdots d\eta_K
\]
Thus 
\begin{align*}
C(\Psi)&=
p(\Psi)/\Psi^{\lambda_1-1}\\
&\propto
\int_0^{\infty} \cdots \int_0^{\infty}
\left\{1+\Psi\left/\prod_{i=2}^K \eta_i\right.\right\}^{-(\lambda_1+c_1)}\left\{\prod_{i=2}^K
\eta_i^{\lambda_i-\lambda_1-1}\left\{1+\eta_i\right\}^{-(\lambda_i+c_i)}\right\}d\eta_2\cdots d\eta_K
\end{align*}
By the dominated convergence theorem

\begin{align*}
\lim_{\Psi\rightarrow 0} C(\Psi)= C(0)\propto
\int_0^{\infty} \cdots \int_0^{\infty}
\left\{\prod_{i=2}^K
\eta_i^{\lambda_i-\lambda_1-1}\left\{1+\eta_i\right\}^{-(\lambda_i+c_i)}\right\}d\eta_2\cdots d\eta_K
\end{align*}
a constant, since we are integrating kernels of GG$(\lambda_i-\lambda_1, \lambda_1+c_i,1)$ distribution and $\lambda_i\geq \lambda_1.$
Therefore, the sparsity parameter of the marginal distribution of $\Psi_i$ is given by the simple form of $\min\{\lambda_i\}.$
\hfill{$\square$}

\subsubsection*{Part (ii)}
Suppose $\Psi_i\sim \GG(\lambda_i,c,d), i=1,2$ then $Y=\Psi_1 + \Psi_2$ has a density 
\begin{eqnarray*} 
f_y(y) & \propto & \int_0^y(y-w)^{\lambda_1-1}\left[1+\frac{(y-w)}{d}\right]^{-(\lambda_1+c)}w^{\lambda_2-1}\left[1+\frac{w}{d}\right]^{-(\lambda_2+c)}dw \\
&=& y^{\lambda_1 +\lambda_2-1} \int_0^1(1-z)^{\lambda_1-1}z^{\lambda_2-1}\left[1+\frac {y(1-z)}{d}\right]^{-(\lambda_1+c)}\left[1+\frac{yz}{d}\right]^{-(\lambda_2+c)}dz \\
&=& y^{\lambda_1 +\lambda_2-1}C(y)
\end{eqnarray*}
and by dominated convergence theorem \\
$\lim_{y\rightarrow 0} C(y) = \int_0^1(1-z)^{\lambda_1-1}z^{\lambda_2-1}dz$
so the sparsity of the convolution is $\lambda_1+\lambda_2.$  This result can be easily generalised to the sum of $K $ independent $\GG(\lambda_i,c,d)$,  $i=1,\ldots,K$  random variables.

\end{document}